\documentclass[final]{cvpr}
\usepackage{times}
\usepackage{epsfig}
\usepackage{graphicx}
\usepackage{amsmath}
\usepackage{amssymb}
\usepackage{comment}
\usepackage{makecell}
\usepackage{soul}
\usepackage{multirow}

\usepackage[pagebackref=true,breaklinks=true,colorlinks,bookmarks=false]{hyperref}

\def\confYear{MICCAI FLARE 2021}

\begin{document}

\title{COBRA: \underline{C}pu-\underline{O}nly a\underline{B}dominal o\underline{R}gan segment\underline{A}tion\\ 
\large A small, fast \& accurate 3D-CNN}

%

\author{Edward G. A. Henderson*
\\
{\tt\small edward.henderson@postgrad.manchester.ac.uk}
\and
D\'{o}nal M. McSweeney*\\
{\tt\small donal.mcsweeney@postgrad.manchester.ac.uk}
\and
Andrew Green\\
{\tt\small andrew.green-2@manchester.ac.uk}
\vspace{0.1in}\\

Division of Cancer Sciences, The University of Manchester\\
Radiotherapy Related Research, The Christie NHS Foundation Trust\\
Manchester, UK\\
{\small *authors contributed equally}
}

\maketitle

\begin{abstract}
Abdominal organ segmentation is a difficult and time-consuming task. To reduce the burden on clinical experts, fully-automated methods are highly desirable. Current approaches are dominated by Convolutional Neural Networks (CNNs) however the computational requirements and the need for large data sets limit their application in practice.\\
\indent By implementing a small and efficient custom 3D CNN, compiling the trained model and optimizing the computational graph: our approach produces high accuracy segmentations (Dice Similarity Coefficient (\%): Liver: $97.3\pm1.3$, Kidneys: $94.8\pm3.6$, Spleen: $96.4\pm3.0$, Pancreas: $80.9\pm10.1$) at a rate of 1.6 seconds per image.\\
\indent Crucially, we are able to perform segmentation inference solely on CPU (no GPU required), thereby facilitating easy and widespread deployment of the model without specialist hardware.
\end{abstract}

\section{Introduction}
Volumetric image segmentation is a time-consuming and often complicated task that requires medical expertise to produce high-quality, reliable delineations. A number of automated approaches have been proposed to free experts from this tedious task and consequently decrease annotation cost. However, variability in acquisition protocols, patient anatomy and the presence of pathologies make this a difficult task to automate. Current state-of-the-art approaches require large and diverse data sets to generalise to unseen examples. The complexity of the resulting models and the size of the CT volumes incur large computational costs and limit applications to centres with adequate computational resources.\\
\indent We aim to improve accessibility to these methods by removing the need for expensive, specialist hardware (GPUs) and instead produce models that can perform quick inference entirely on CPU. To reduce the computational cost of image processing, high-resolution 3D CT scans are all downsampled to the same size. Our model architecture is inspired by the 3D U-Net presented in \cite{cicek2016} with a few notable modifications to reduce model size and the number of FLOPs (discussed in section \ref{sec:Method}). Finally, we use the Open Neural Network Exchange (ONNX) \cite{bai2019} to compile the trained model, at which point we apply further optimisation to the computational graph.\\
\indent As a result, we are able to deploy a small (1.7 Mb) and fast model (1.6 seconds/image on CPU) that produces high-quality 3D segmentations of abdominal organs.

\section{Method}
\label{sec:Method}
\begin{figure*}[ht]
\centering
\includegraphics[width=\textwidth]{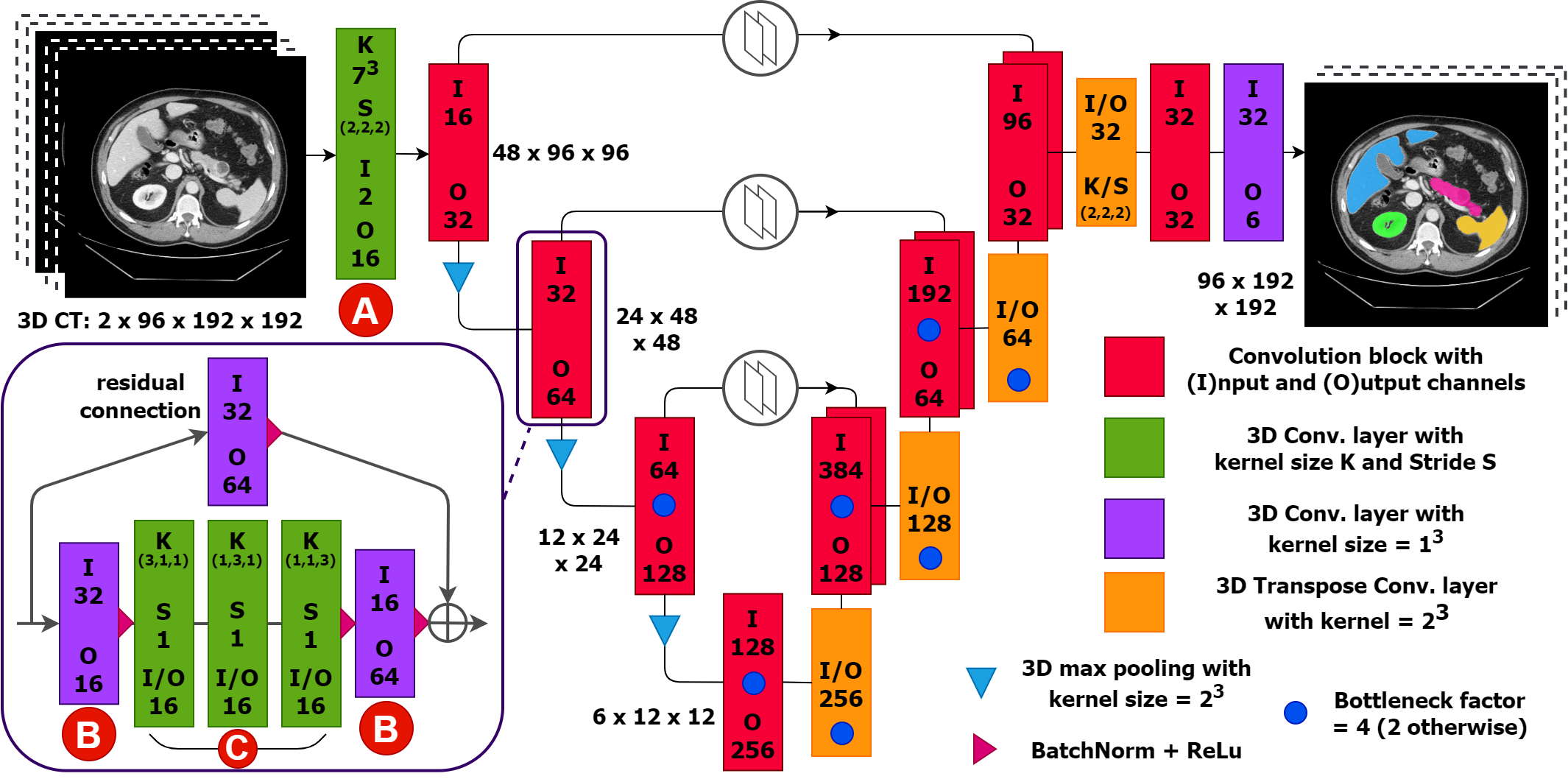}
\caption{A diagram of our custom 3D CNN architecture. Our network is influenced by a standard 3D UNet \cite{cicek2016} with added ResNet-like residual connections \cite{he2016}. In order to reduce the model size and computational load we introduce three components: a YOLO-inspired \cite{redmon2016} 7x7x7 input convolution with stride 2 to quickly reduce the size of the input image whilst preserving a large field of view (see A); bottleneck structures to reduce the number of kernels required (B); and asymmetric factorisation of convolution layers (C).}
\label{fig:Network}
\end{figure*}

Our solution for this challenge is based on a single CNN model which performs segmentation inference on an entire downsampled CT scan. We developed a custom 3D CNN for this challenge which is illustrated in Figure~\ref{fig:Network}.

\subsection{Preprocessing}
When training, we downsample all CT scans and gold standard segmentations to a standardised resolution of $96\times192\times192$. The CT scans are downsampled using 3\textsuperscript{rd} order spline interpolation, whereas nearest-neighbour downsampling is used for the corresponding gold standard segmentations. No cropping is applied prior to training. These decisions were made to promote segmentation scale invariance within our model.

The liver, kidneys, spleen and pancreas are all soft tissue structures. Therefore, we chose to normalise the CT scans prior to training using windowing (grey-level mapping). By using windowing, we can enhance the contrast of these soft tissue structures whilst also mapping the voxel intensities onto the range $[0,1]$ to improve learning stability. The CT scans are normalised in two separate contrast channels which are concatenated (see the input in Figure \ref{fig:Network}). This improved our segmentation performance. In channel one, we use a contrast setting which is used to view general soft tissue structures in the abdominal area (W400, L50). In the second channel we apply a tighter window (W100, L60) in an attempt to increase the contrast of the pancreas.

To aid in the determination of organ boundaries, we split the background label (0) in two: ``air'' and ``body''; this is done by thresholding at -200~HU and applying binary closing and hole filling operations to the resulting mask. Air retains the original background label 0, and body becomes label 1, with all other labels shifted accordingly. 


\begin{table*}[!htbp]
\caption{Data splits of FLARE2021.}
\label{tab:dataset}
\centering
\begin{tabular}{llll}
\hline
Data Split  
& Center & Phase & \# Num.\\
\hline
\multirow{2}{*}{Training ( 361 cases )} & The National Institutes of Health Clinical Center& portal venous phase & 80  \\
    & Memorial Sloan Kettering Cancer Center    & portal venous phase& 281 \\
\hline
\multirow{3}{*}{Validation ( 50 cases )}  & Memorial Sloan Kettering Cancer Center    & portal venous phase& 5   \\
    & University of Minnesota & late arterial phase& 25  \\
    & 7 Medical Centers& various phases& 20  \\
\hline
\multirow{4}{*}{Testing ( 100 cases )} & Memorial Sloan Kettering Cancer Center    & portal venous phase& 5   \\
    & University of Minnesota & late arterial phase& 25  \\
    & 7 Medical Centers & various phases& 20  \\
    & Nanjing University   & various phases& 50\\
\hline
\end{tabular}
\end{table*}

\subsection{Proposed Method}
In Figure~\ref{fig:Network} we show our custom 3D CNN designed for this challenge. The architecture is inspired by the classic 3D UNet \cite{cicek2016} with added residual connections \cite{he2016}. Residual connections are now a standard addition to most modern 3D UNet CNNs and improve gradient propagation in the training process. 

Three key features of our model, which reduce the model size (number of parameters) and computational complexity (number of FLOPs), are introduced below. 

First, a YOLO-inspired \cite{redmon2016} input layer is added, consisting of a 7x7x7 convolution with stride of 2. In Figure~\ref{fig:Network} we mark this particular layer `A'. This input layer quickly reduces the size of the volume being processed by the network, while preserving a wide field-of-view. By reducing the size of the CT scan being processed by the network, segmentation inference is accelerated.

To compress our model, we implemented bottleneck structures as used by He et al.~for their deeper ResNet architectures \cite{he2016}. Bottlenecks are introduced to reduce and then restore the number of kernels in convolutional layers in certain portions of the model. By compressing the number kernels of convolutional layers deep in the UNet architecture, we force our model to learn better feature representations whilst also reducing its size and computational complexity. 

Bottlenecks are constructed by sandwiching existing convolutional layers with 1x1x1 convolutions which first reduce and then restore the number of kernels. An example bottleneck structure is labelled `B' Figure \ref{fig:Network}. In our model we implement bottlenecks around all the traditional encoder and decoder double convolution layers (shown in red in Figure \ref{fig:Network}). In addition, all 3D transpose convolutions are similarly bottlenecked. For each bottleneck, we define a ``bottleneck factor" which determines the multiplicative reduction in kernels required. By default, we use a bottleneck factor = 2, except in blocks marked with a small blue dot in figure \ref{fig:Network} where we use a bottleneck factor = 4. This higher level of kernel compression is performed in the widest layers of the CNN to further reduce model size and accelerate inference.

To further compress the size of the model and reduce the computational complexity we applied 3D asymmetric factorisation to the convolutional layers of our model. Factorisation of convolutional kernels was introduced by Szegedy et al. for their Inception v2 architecture \cite{Szegedy2016a} and has been more recently applied to 3D data by Yang et al. for action recognition in consecutive video frames \cite{Yang2019}. Asymmetric factorisation simulates a large 3D kernel using a series of 1D kernels. For example, we factorise all 3x3x3 convolution kernels into a series of layers with 3x1x1, 1x3x1 and 1x1x3 kernels. A factorised 3x3x3 kernel reduces the number of parameters by up to a third compared to a standard layer. Additionally, the number of FLOPs required to perform factorised layers is significantly lower. For our final model, we applied asymmetric factorisation to all convolutional layers with 3x3x3 kernels and the YOLO-inspired 7x7x7 convolutional layer. Please refer to the section marked `C' in Figure \ref{fig:Network} which shows the layers resulting from asymmetric factorisation.

As a result, our custom 3D CNN contains just 436,982 parameters and requires 48 GFLOPs to segment a $96\times192\times192$ CT volume.


\subsection{Post-processing}
Our CNN model outputs raw segmentation predictions at a resolution of $96\times192\times192$. We apply \textit{argmax} to obtain the hard predictions before upsampling the segmentations to the original CT scan dimensions using nearest-neighbour upsampling.

\section{Dataset and Evaluation Metrics}
\subsection{Dataset}
\label{ssec:dataset}
\begin{itemize}
\item The dataset used for FLARE2021 is adapted from MSD~\cite{simpson2019MSD} (Liver~\cite{bilic2019lits}, Spleen, Pancreas), NIH Pancreas~\cite{roth9data,roth2015deeporgan,clark2013cancer}, KiTS~\cite{KiTS,KiTSDataset}, and Nanjing University under the license permission. For more detail information of the dataset, please refer to the challenge website and~\cite{Ma-2021-AbdomenCT-1K}.

\item Details of training / validation / testing splits:\\
The total number of cases is 511. An approximate 70\%/10\%/20\% train/validation/testing split is employed
resulting in 361 training cases, 50 validation cases, and 100 testing cases. Detailed information is presented in Table~\ref{tab:dataset}.
\end{itemize}

At this stage we only have access to the training subset of images and their corresponding gold standard segmentations. We present a 5-fold cross validation of our model for this subset in anticipation of evaluation of our model with the validation and testing subsets.

\subsection{Evaluation Metrics} 
\begin{itemize}
    \item Dice Similarity Coefficient (DSC)
    \item Normalized Surface Distance (NSD) with tolerance of 1mm
    \item Running time: $\sim1.6$ seconds per CT scan
    \item Maximum used GPU memory - 0 Mb
\end{itemize}

\begin{table}[!htbp]
\caption{Environments and requirements.}
\label{table:env}
\begin{center}
\resizebox{0.47\textwidth}{!}{
\begin{tabular}{m{2.5cm}<\raggedright|m{6cm}<\raggedright} 
\hline
Windows/Ubuntu version & Ubuntu 20.04.2 LTS \\
\hline
CPU & AMD Ryzen 9 3950X 16-Core Processor \\
\hline
RAM & 64GB \\
\hline
GPU & NVidia GeForce RTX 3090 \\
\hline
CUDA version & 11.1 \\ 
\hline
Programming language & Python 3.8 \\ 
\hline
Deep learning framework & Pytorch (Torch 1.8.1, torchvision 0.9.1) \\
\hline
Specification of dependencies & SimpleITK (2.02), onnx (1.9.0), onnxruntime (1.8.1), numba (0.53.1), numpy (1.21.1) \\
\hline
(Optional) code is publicly available at & https://github.com/rrr-uom-projects/FLARE21 \\
& Docker image: afgreen/flare21:latest (on dockerhub) \\
\hline
\end{tabular}
}
\end{center}
\end{table}

\begin{table}[!htbp]
\caption{Training protocols.}
\label{table:training}
\begin{center}
\resizebox{0.47\textwidth}{!}{
\begin{tabular}{m{2.5cm}<\raggedright|m{6cm}<\raggedright} 
\hline
\makecell*[c]{Data augmentation\\ methods} & Shifting ($\pm 4$ voxels, 	$\sim10$mm), rotations (in-plane: $\pm 10^{\circ}$), scaling (80\% - 120\%)\\
\hline
\makecell*[c]{Initialization of\\ the network} & ``He" normal initialization \cite{He2015a}\\
\hline
Patch sampling strategy & None - Full CT downsampled and used as input \\
\hline
Batch size & 4 \\
\hline 
Patch size & 96$\times$192$\times$192  \\ 
\hline
Maximum epochs & 1000 \\
\hline
Optimizer     & Adam     \\ \hline
Initial learning rate  & 0.001 \\ \hline
Learning rate decay schedule & Reduce LR on plateau with patience=75 epochs \\
\hline
Stopping criteria, and optimal model selection criteria & Early stopping when the validation loss does not improve for 175 epochs. Optimal model is chosen based on best validation loss.\\
\hline
Loss function & Weighted multi-class soft Dice \\
\hline
Training time & 10-15 hours \\
\hline
\end{tabular}
}
\end{center}
\end{table}

\section{Implementation Details}

\subsection{Environments and requirements}
The environments and requirements of our method are shown in Table~\ref{table:env}.

\begin{table*}[htbp]
\normalsize
\centering
\caption{Quantitative results of 5-fold cross validation in terms of DSC and NSD. We show the median and standard deviation of both measures for every fold individually and for all training folds combined (361 images).}
\label{tab:cross-validation}
\renewcommand\tabcolsep{3pt}
\begin{tabular}{cccccccccc}
\hline
\multirow{2}{*}{Training}   & \multicolumn{2}{c}{Liver}  & \multicolumn{2}{c}{Kidney}  & \multicolumn{2}{c}{Spleen}  & \multicolumn{2}{c}{Pancreas}                                       \\
\cline{2-9}   & DSC (\%)   & NSD (\%) & DSC (\%)   & NSD (\%) & DSC (\%)   & NSD (\%) & DSC (\%)   & NSD (\%)  \\
\hline
Fold 1 & 97.3$\pm$ 0.9 & 85.7$\pm$ 4.5 & 94.8$\pm$ 2.2 & 87.1$\pm$ 5.5 & 96.4$\pm$ 2.0 & 91.7$\pm$ 5.3 & 81.5$\pm$ 9.0 & 55.5$\pm$ 11.9 \\ 
\hline
Fold 2  & 97.3$\pm$ 2.1 & 85.8$\pm$ 7.0 & 94.9$\pm$ 4.9 & 88.3$\pm$ 8.2 & 96.5$\pm$ 3.1 & 92.0$\pm$ 6.5 & 81.2$\pm$ 7.3 & 59.9$\pm$ 10.2 \\
\hline
Fold 3 &  97.1$\pm$ 1.5 & 84.8$\pm$ 5.9 & 94.6$\pm$ 4.3 & 86.3$\pm$ 7.0 & 96.2$\pm$ 2.7 & 91.9$\pm$ 5.3 & 80.1$\pm$ 13.2 & 54.5$\pm$ 12.6 \\
\hline
Fold 4 & 97.1$\pm$ 0.8 & 85.6$\pm$ 5.1 & 94.8$\pm$ 3.6 & 87.0$\pm$ 5.8 & 96.3$\pm$ 1.3 & 92.1$\pm$ 4.5 & 81.0$\pm$ 9.3 & 55.9$\pm$ 13.7 \\
\hline
Fold 5  & 97.4$\pm$ 1.0 & 86.7$\pm$ 4.9 & 94.9$\pm$ 1.5 & 88.4$\pm$ 5.4 & 96.3$\pm$ 4.7 & 92.3$\pm$ 8.7 & 80.0$\pm$ 10.4 & 57.6$\pm$ 12.3 \\
\hline
Median & 97.3$\pm$ 1.3 & 85.9$\pm$ 5.6 & 94.8$\pm$ 3.6 & 87.5$\pm$ 6.5 & 96.4$\pm$ 3.0 & 92.0$\pm$ 6.3 & 80.9$\pm$ 10.1 & 56.8$\pm$ 12.3 \\
\hline
\end{tabular}
\end{table*}
    
\subsection{Training protocols}
The training protocols we used to train our custom 3D CNN are shown in Table~\ref{table:training}. We conducted a five-fold cross validation using the ``Training" subset outlined in section \ref{ssec:dataset} containing 361 CT scans. For the cross-validation, each set of training, validation and test folds contained 252, 36 and 73 images respectively.

\begin{figure}[htpb]
\centering
\includegraphics[scale=0.45]{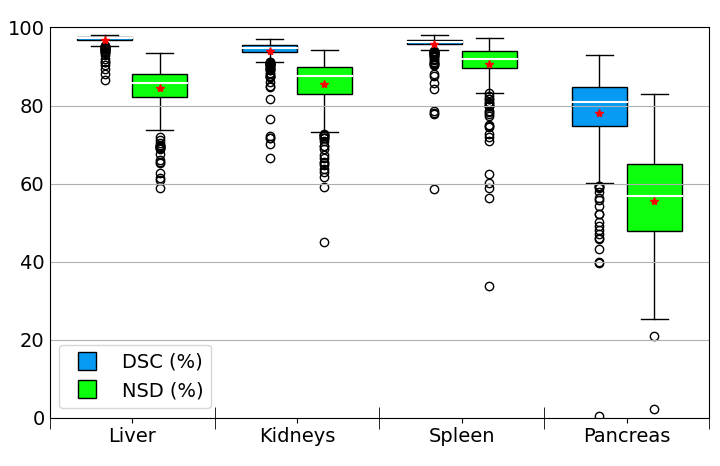}
\caption{Boxplots of the organ segmentation results (DSC and NSD) of the 5-fold cross validation. The white line shoes the median and red star the mean.}
\label{fig:5-fold}
\end{figure}

\subsection{Testing protocols}
In the testing phase, inference is performed on the entire CT volume by downsampling the input image to a size of $96\times192\times192$ with 3rd order spline interpolation. A Gaussian kernel was applied in-plane to prevent aliasing artefacts when downsampling. As in the training phase, we do not crop images prior to inference.

\indent We use ONNX \cite{bai2019} to compile the trained model thereby decreasing model size - enabling inference on CPU - and increasing inference speed. At this point, a number of graph level transformations are applied to improve model performance further. These include: constant folding, redundant node elimination and node fusion. This led to a small but noticeable improvement in inference speed.\\
\indent We experimented with dynamic and static quantization. Though both reduced model size (1.7Mb $\rightarrow$ 614 Kb), the former increased inference time by an order of magnitude while the latter dramatically reduced model performance. Therefore, we chose not to apply weight quantization. Quantization-aware training may be an alternative for future work.\\
\indent Following inference, \textit{argmax} is applied to model predictions to convert them to multi-class masks then, they are upsampled to the original scan dimensions with nearest neighbour interpolation. \\

\section{Results}
\subsection{Quantitative results for 5-fold cross validation.}
The provided results are based on the 5-fold cross-validation results and validation cases.
Table~\ref{tab:cross-validation} illustrates the results of 5-fold cross-validation. Figure~\ref{fig:5-fold} is the corresponding boxplot of organ segmentation performance. While high DSC and NSD scores are obtained for the liver, kidney and spleen, accuracy is lower for the pancreas - highlighting the difficulty of segmenting this organ.

Fold 2 was our highest performing model in terms of DSC and NSD across the four segmented organs, so we selected this model to submit for evaluation with the validation and test sets.

\subsection{Quantitative results on validation set.}
 Table~\ref{tab:quanti-validation} illustrates the results on validation cases. Comparison between Table~\ref{tab:cross-validation} and Table~\ref{tab:quanti-validation} illustrates better DSC and NSD performance is obtained for the 5-fold cross validation than the validation set.

\begin{table}[!htbp]
\caption{Quantitative results on validation set.}
\label{tab:quanti-validation}
\centering
\begin{tabular}{ccc}
\hline
Organ    & DSC (\%)       & NSD (\%)        \\
\hline
Liver    & 90.6$\pm$13.5 & 62.7$\pm$18.2  \\
Kidney   & 68.1$\pm$29.5  & 54.8$\pm$25.9  \\
Spleen   & 83.8$\pm$21.6  & 66.1$\pm$21.3  \\
Pancreas & 53.9$\pm$26.6  & 38.2$\pm$21.7  \\
\hline
\end{tabular}
\end{table}

\subsection{Qualitative results}
\begin{figure}[ht]
\centering
\includegraphics[width=\columnwidth]{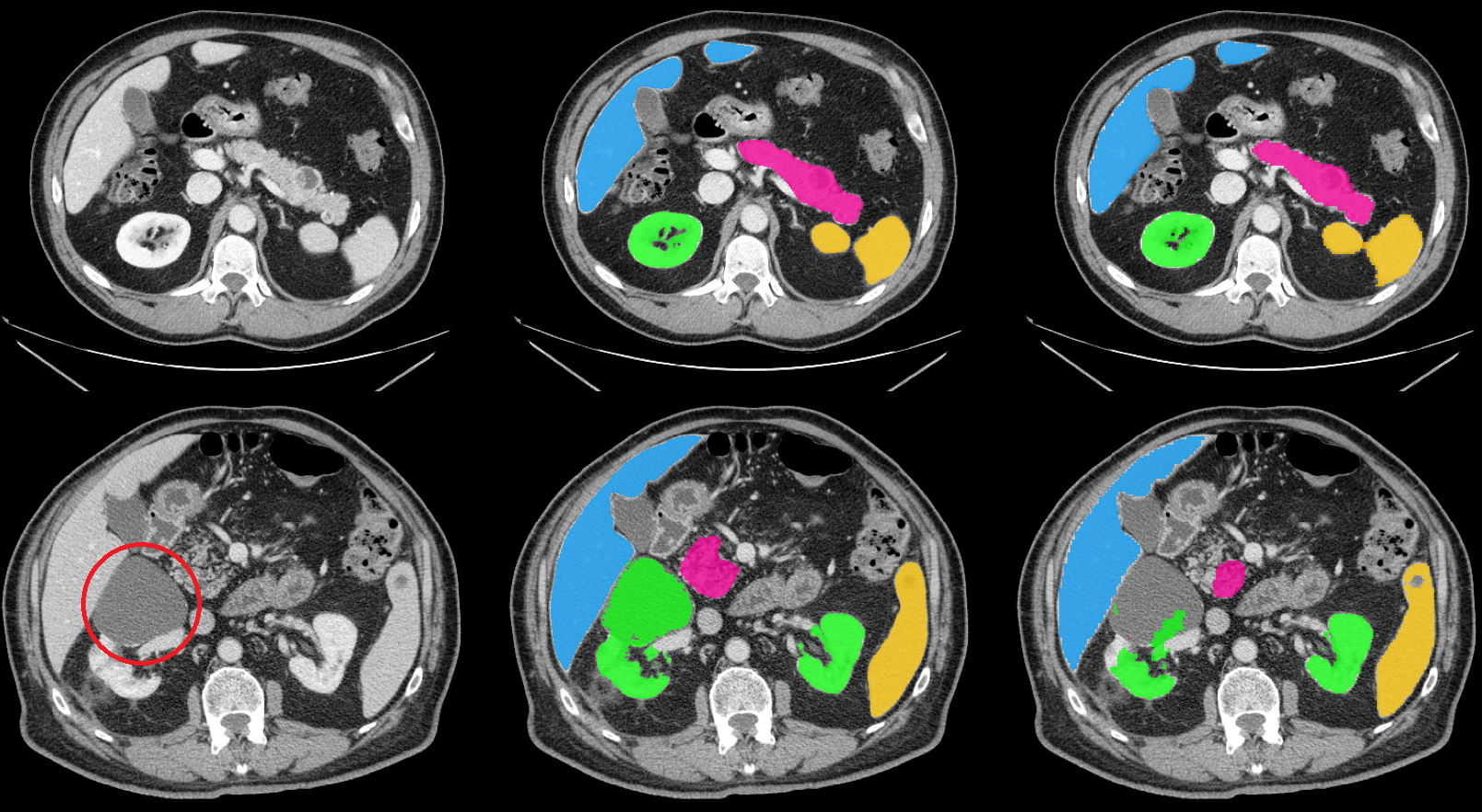}
\caption{Two sets of axial slices of segmentation outputs from our model. Both patients shown here are from the training subset of images. In the first column is the CT image alone, the second column shows the gold standard segmentations and the third shows our models prediction. In the top row we show an example where our model performs well, accurately segmenting the liver (blue), kidneys (green), pancreas (pink) and spleen (yellow) with good accuracy. In the bottom row we show an example where our model struggles to segment the right kidney. In this example a tumour is present in the right kidney which our model is unable to recognise.}
\label{fig:quanli}
\end{figure}

In Figure \ref{fig:quanli} we present two examples of segmentations from the training subset predicted by our model. In the top row we show an axial slice of a patient where our model performs well. From left to right we show the CT scan alone, the gold-standard segmentations and the organ segmentation predictions made by our model. In this example, our model segments the liver (\textit{blue}), kidneys (\textit{green}), pancreas (\textit{pink}) and spleen (\textit{yellow}) with high accuracy compared to the gold standard. Our model struggles in resolving the internal structure of the right kidney, where the gold standard segmentation has avoided the calyx. In the bottom row, we show an example where our model struggles to produce an accurate segmentation for the kidneys (\textit{green}). In this example, a large tumour (circled in red) has infiltrated the right kidney. In the gold standard this tumour is included in the kidney segmentation, however, our model fails to classify this structure as part of the kidney.

\begin{figure*}[ht]
\centering
\includegraphics[width=\textwidth]{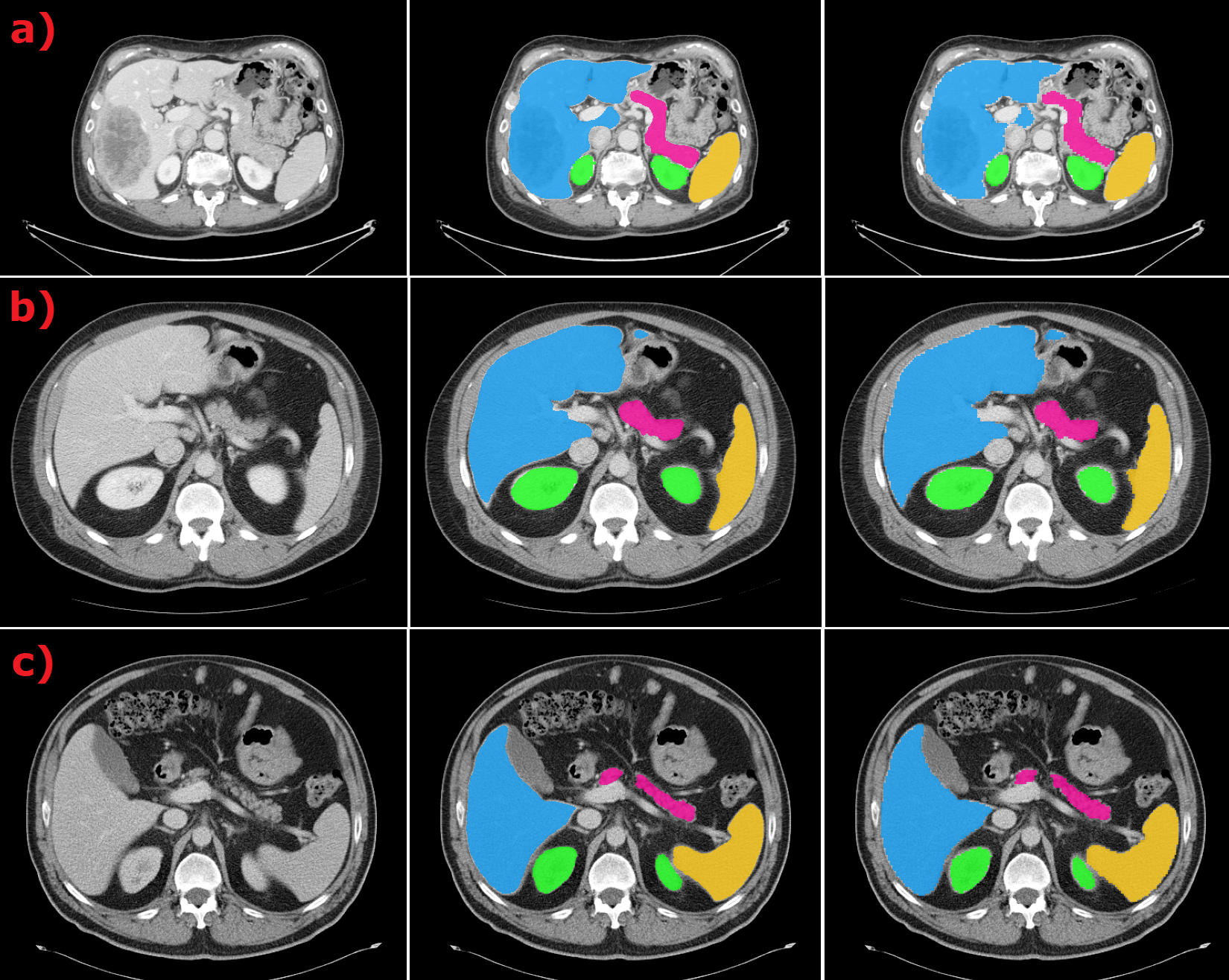}
\caption{Three examples from the validation set where our model performs very well. As in figure~\ref{fig:quanli}, the first column shows the CT, the second the golden standard and the third shows our model's prediction. Our model can segment the abdominal organs in a diverse range of shapes and sizes reflecting common anatomical variation.}
\label{fig:val_good_ims}
\end{figure*}

In figure~\ref{fig:val_good_ims} we show three examples from the validation set where our model does a great job at segmenting the organs. These examples show the degree of anatomical variation commonly observed for these abdominal organs. The segmentation predictions made by our model (right column) closely reflect the gold standard (central column).

\begin{figure*}[ht]
\centering
\includegraphics[width=\textwidth]{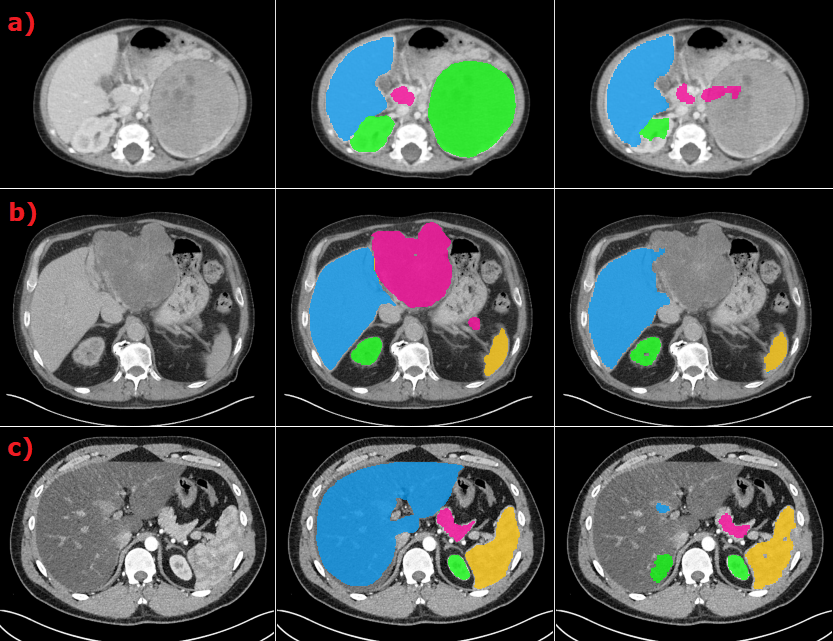}
\caption{Three challenging examples drawn from the validation set where there is significant anatomical deformation caused by disease. As in figure~\ref{fig:quanli}, the first column shows the CT, the second the golden standard and the third shows our model's prediction. Our CNN is unable to segment the left kidney in case a) which is much larger than normal, likely infiltrated by a tumour. Similarly, our model is unable to segment the pancreas of case b) which is larger than standard and oddly shaped. The liver of case c) is darker than normal, often indicative of fatty liver disease. Due to it's darker appearance, our model is unable to segment this liver.}
\label{fig:val_poor_ims}
\end{figure*}

In figure~\ref{fig:val_poor_ims} we show three challenging examples from the validation subset. In each of these cases, disease is present which abnormally alters the shape, size or appearance of one or more of the organs. In row a), the left kidney (green) is abnormally large, possibly infiltrated by a tumour, and is has been failed to be segmented by our model. In row b), the pancreas (pink) is atypically enlarged and shaped. Our CNN is unable to segment the pancreas in this case. Finally, in row c) the liver (blue) is darker than usual. This is often observed in fatty liver cases. Our model is unable to segment this liver, even though the liver is still easily visible. These three cases are uncommon and it is likely that our model was not exposed to many similar examples in the training phase.

\section{Discussion and Conclusion}
We have developed a method for 3D segmentation of organs in abdominal CTs which is capable of producing high quality liver, kidney, spleen and pancreas segmentations in 1.6 seconds using only the CPU.

Medical image segmentation with CNNs is now very common and a well-studied field. However, many methods which produce accurate segmentations very quickly on a GPU will suffer significant slowdown when operating on a CPU. For example, Panda et al. \cite{panda2021} developed a 3D auto-segmentation method which produces high-quality segmentations of the pancreas (DSC = 0.91), but takes 4 minutes to perform inference for a single image on CPU.

We have performed a 5-fold cross-validation using the training subset of images (361 images). Our model performance was consistent across all five folds. However, when evaluated on the validation set, our models segmentation performance was substantially lower in all organs. This phenomenon may caused by over-fitting the model on the training set. However, we believe the significant deterioration in segmentation performance is due to the distribution of the contrast phases of images between the data splits. Our model was trained only with portal venous phase images, whereas at least half of the validation set is made up of images from an earlier contrast stage (late arterial). 

Since we are unable to use external data for this challenge, in future we plan to include additional data augmentation to simulate different contrast phases in the training stage. In addition, we would like to ensure that any future training datasets contains examples of fatty livers and other diseases to ensure the model is robust to cases similar to the ones shown in figure~\ref{fig:val_poor_ims}.

One of the CTs in the training subset (\textit{train\_270\_0000.nii.gz}) was missing many slices of the image, including large portions containing the liver and spleen. As a result, we excluded this image from the calculation of DSC and NSD.

Automated segmentation of medical images with CNNs is now the state-of-the-art \cite{Cardenas2019}, and will increasingly release clinicians from the time-intensive task of manually annotated structures. Our model performs fast and accurate 3D segmentation on the CPU, which enables much wider deployment of such models within clinics and research groups since it does not require the use of specialist hardware (GPUs).


\newpage

\section*{Acknowledgement}
The authors of this paper declare that the segmentation method they implemented for participation in the FLARE challenge has not used any pre-trained models nor additional datasets other than those provided by the organizers.

{\small
\bibliographystyle{IEEEtran} 
\bibliography{egbib}
}

\end{document}